\documentclass[aps,prl,twocolumn,superscriptaddress,showpacs,showkeys]{revtex4}

\usepackage{graphicx,amsmath}
\usepackage{times}

\begin{document}

\newcommand{\eg}{{\it e.g.}}
\newcommand{\ie}{{\it i.e.}}
\newcommand{\pc}{p_{\rm c}}

\newcommand{\kav}{\left<k\right>}

\newcommand{\affila}{
 State Key Laboratory of Theoretical Physics, Institute of Theoretical Physics, Chinese Academy of Sciences, P.O. Box 2735, Beijing 100190, China
}

\title{General Clique Percolation in Network Evolution}

\author{Jingfang Fan }
\affiliation{\affila}
\author{Xiaosong Chen}
\email{chenxs@itp.ac.cn}
\affiliation{\affila}

%\date[]{\protect\today}
\date[]{}

\begin{abstract}
We introduce a general $(k,l)$ clique community, which consists of adjacent $k$-cliques sharing at least $l$ vertices with $k-1 \ge l \ge 1$.
The emergence of a giant $(k,l)$ clique community indicates a $(k,l)$ clique percolation, which is studied by the largest size gap $\Delta$
of the largest clique community during network evolution and the corresponding evolution step $T_c$. For a clique percolation, the averages
of $\Delta$ and $T_c$ and the root-mean-squares of their fluctuations have power law finite-size effects whose exponents are related to the
critical exponents. The fluctuation distribution functions of $\Delta$ and $T_c$ follow a finite-size scaling form. In the evolution of
the Erd\H{o}s-R\'enyi network, there are a series of $(k,l)$ clique percolation with $(k,l)=(2,1),(3,1),(3,2),(4,1),(4,2),(5,1),(4,3)$, and so on.
The critical exponents of clique percolation depend on $l$, but are independent of $k$. The universality class of a $(k,l)$ clique percolation
is characterized alone by $l$.
\end{abstract}


%02.10.Ox Combinatorics; graph theory
%89.75.Hc Networks and genealogical trees
%05.70.Fh Phase transitions: general studies
%64.60.-i General studies of phase transitions

\maketitle

Community structure is of great interest in the studies of networks \cite{gn-pnas,fortunato2010community}. The term network community is defined as a group
of vertices that are more densely connected each other than other vertices in a network.  The clique \cite{bollobas-book} and core \cite{Dorogovtsev1}
are the two examples of network community. To analyze the overlapping community structure of networks, Palla \emph{et al.}\cite{Pallasubm} proposed the
clique percolation method (CMP) to build up the communities from $k$-cliques, which is a fully connected subset of $k$ vertices. Two $k$-cliques are
considered to be adjacent if they share $k-1$ vertices. A clique community is defined as the maximal union of $k$-cliques that can be reached from each
through a series of adjacent $k$-cliques.

When the size of a clique community is comparable to network size $N$, a giant clique community emerges and there is a clique percolation.
Der{\'e}nyi \emph{et al.} \cite{Derenyi} studied the clique percolation of the Erd\H{o}s-R\'enyi (ER)
model\cite{e-r}. The $k$-cliques percolation takes place when the probability of connecting two vertices in the network reaches the
threshold $\pc(k)=[(k-1)N]^{-1/(k-1)}$ \cite{Derenyi}. The normal percolation transition of ER model corresponds to the $k=2$ clique percolation.

\begin{figure}[b!]
\centerline{\includegraphics[angle=0,width=0.85\columnwidth]{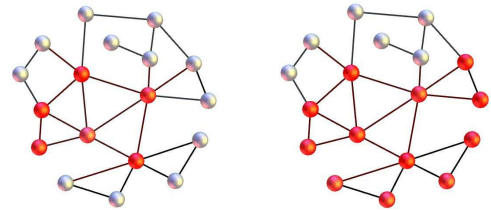}}
\caption{(color online)
Sketches of different clique communities in a network. The vertices belonging to a clique community are marked with red color.
The largest $(3,2)$ clique community is shown on the left and the largest $(3,1)$ clique community on the right.}
\label{fig_sketch}
\end{figure}

In this Letter, we introduce a general $(k,l)$ clique community, where any two adjacent $k$-cliques share at least $l$ vertices with $k-1 \ge l \ge 1$.
For the same network, the clique communities of different $(k,l)$ are different. As an illustration, $(3,1)$ and $(3,2)$ clique
communities in a network are shown in Fig.\ref{fig_sketch} and they are different. A $(k,l)$ clique percolation appears with the emergence of a giant $(k,l)$ clique community. The previous clique percolation discussed in Ref.\cite{Derenyi} corresponds to the special case $l=k-1$ of general $(k,l)$ clique percolation. The normal percolation transition corresponds to a $(2,1)$ clique percolation. Here we study the general $(k,l)$ clique percolation of the ER model by analyzing the finite-size effects of network evolution. From the power law exponents of finite-size effects, we can obtain the critical exponents of $(k,l)$ clique percolation.

At first, we calculate the threshold of the $(k,l)$ clique percolation using the approach developed by Newman, Watts, and Strogatz\cite{newman-arbitPk}.
From the probability distribution $p_k$ of vertex degrees, we define the generating functions $G_0(x) = \sum_{k=0}^\infty p_k x^k$ and
$G_1(x) = \sum_{k=0}^\infty q_k x^k$, where $q_k=p_{k+1}(k+1)/<k>$ is the excess degree distribution of network \cite{newman2010networks}.

The generating function $H_1(x)$ of the probability distribution for the sizes of the components reached from a randomly chosen edge satisfies
a self-consistent equation
\begin{equation}
H_1(x) = x G_1(H_1(x)).
\label{3}
\end{equation}
When its first derivative $H'_1 (1)$ becomes infinite, a giant component appears and there is a percolation. This corresponds to $z'=G'_1(1) = 1$.

After taking $k$-clique as the new unit and considering two cliques to be adjacent if they share at least $l$ vertices, the average degree
of a $k$-cliques is
\begin{equation}
z  = \sum_{k'=l}^{k-1} \binom{N-k}{k-k'}\binom{k}{k'}p^{\binom{k}{2} - \binom{k'}{2}},
\label{5}
\end{equation}
where $p$ is the probability to connect two vertices with an edge. Correspondingly, the average excess degree of the $k$-cliques is
\begin{equation}
z'  = \sum_{k'=l}^{k-1} \binom{N-k}{k-k'}\left[\binom{k}{k'}-1\right]p^{\binom{k}{2} - \binom{k'}{2}}.
\label{51}
\end{equation}

The threshold $p_c(k,l)$ of $(k,l)$ clique percolation can be determined from the equation
\begin{equation}
\sum_{k'=l}^{k-1} \frac{\binom{k}{k'}-1}{(k-k')!}\left[1+O(N^{-1})\right]\left[N p_c^{\frac{k+k'-1}{2}}\right]^{k-k'}=1\;,
\end{equation}
which gives
\begin{eqnarray}\label{8}
p_c (k,l) &=& a (k,l) N^{- \frac{2}{k+l-1}}\left[1+O(N^{-\frac{k-l-1}{k+l-1}})\right], \\
a (k,l) &=& \left[\frac{\binom{k}{l}-1}{(k-l)! } \right]^{-\frac{2}{(k-l)(k+l-1)}}\;.
\end{eqnarray}
For $l=k-1$, the correction term in Eq.\ref{8} vanishes and $\pc(k,k-1)=[(k-1)N]^{-\frac{1}{k-1}}$, which is in agreement with the result of Der{\'e}nyi
et al \cite{Derenyi,palla2007critica}. For $l < k-1$, the correction term $O(N^{-\frac{k-l-1}{k+l-1}})$ exists. For general $(k,l)$, only $p_c (k,l) \sim
N^{-\frac{2}{k+l-1}}$ is obtained by Bollob{\'a}s and Riordan\cite{bollobas}.

\begin{figure}[t!]
\centerline{\includegraphics[angle=0,width=0.75\columnwidth]{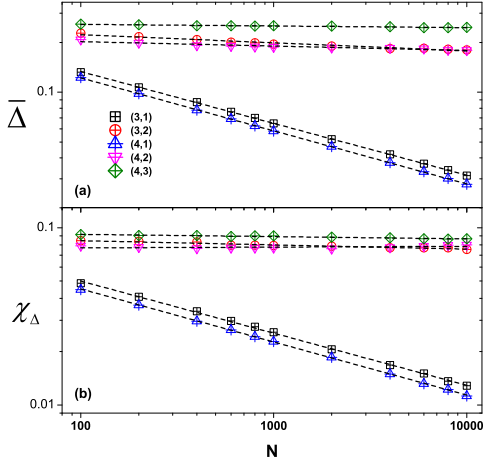}}
\caption{(color online) Log-log plot of the average size gap $\bar{\Delta}$ and the root-mean-square of its fluctuation $\chi_{\Delta}$
versus $N$ for different $(k,l)$ clique percolation. Power law behaviors of $\bar{\Delta}$ and $\chi_{\Delta}$ are confirmed by
simulation data. From the slopes of fitting lines, $\beta_1$ and $\beta_2$ can be obtained and are summarized in Table I.
}
\label{fig_phi}
\end{figure}

In the following, we study $(k,l)$ clique percolation of ER model with Monte Carlo simulation and finite-size effects of network
evolution \cite{fanjingfang}. During an evolution process, the largest $(k,l)$ clique community has a size gap $S_1 (T)-S_1 (T-1)$
at an evolution step $T$. The largest reduced size gap during the whole evolution process is
\begin{equation}
\Delta \equiv \frac{1}{N}max\left(\{S_1(T)-S_1(T-1)\}\right),
\end{equation}
which could be related to a percolation transition at the corresponding evolution step $T_c$. From the results of $\Delta$ and $T_c$ in many Monte Carlo simulations, we can calculate the average size gap $\bar{\Delta}$ and the average transition point $\bar{T}_c$. Since $T=p\times N(N-1)/2$ and therefore $\bar{T}_c \propto N^{2-\frac{2}{k+l-1}}$, it is convenient to introduce a reduced evolution step $r=T/N^{2-\frac{2}{k+l-1}}$ and the reduced transition point $r_c \equiv T_c/N^{2-\frac{2}{k+l-1}}$. We anticipate that $\bar{\Delta}$ and $\bar{r}_c$ have the power law finite-size effects as
\begin{eqnarray}
\label{9d}
\bar{\Delta}(N) &\sim& N^{-\beta_{1}},\\
 \bar{r}_c (N)-r_c (\infty) &\sim& N^{-1/\nu_1}.
\label{9}
\end{eqnarray}
Using $p_c (k,l)$ in Eq.~\ref{8}, we get
\begin{equation}
\label{9c}
r^{T}_c(\infty) = {1\over 2}a(k,l).
\end{equation}
The character of clique percolation is determined by the exponent $\beta_1$. The clique percolation is continuous when $0< \beta_1 < 1$ and discontinuous when $\beta_1=0$ \cite{Nagler}.

The fluctuations $\delta \Delta = \Delta -\bar{\Delta}(N)$ and $\delta r_c = r_c - \bar{r}_c (N)$ are investigated also. Their root-mean-squares are defined as
\begin{eqnarray}
\chi_\Delta &=& \sqrt{\left<[\delta \Delta]^2\right>},\\
\chi_r &=& \sqrt{\left< [\delta r_c]^2\right>},
\label{10}
\end{eqnarray}
which decay algebraically as
\begin{eqnarray}
\label{11}
\chi_{\Delta} & \sim &  N^{-\beta_2},\\
\chi_{r} & \sim &  N^{-1/\nu_2}.
\label{12}
\end{eqnarray}

We anticipate a finite-size scaling form of fluctuation distribution functions as
\begin{eqnarray}\label{13}
P_\Delta(\delta \Delta,N) & = & N^{\beta_{2}}f_{1}(\delta \Delta N^{\beta_2}),\\
P_r(\delta r_c,N) & = & N^{1/\nu_{2}}f_{2}(\delta r_c N^{1/\nu_2}).
\label{14}
\end{eqnarray}

The universality class of continuous clique percolation is characterized by the critical exponents $\beta_1$, $\beta_2$, $\nu_1$ and $\nu_2$.
Different clique percolation with the same critical exponents belong to the same universality class.

In Fig.\ref{fig_phi}(a), the average $\bar{\Delta}$ is plotted with respect to network size $N$ for $(k,l)=(3,1), (3,2), (4,1), (4,2)$, and $(4,3)$.
The log-log plot of $\bar{\Delta}$ versus $N$ show that $\bar{\Delta} \propto N^{-\beta_1}$. From the slope of fitting line, we can get the exponent
$\beta_1$. It has been obtained that $\beta_1 = 0.331(5)$ for $(2,1)$, $\beta_1 = 0.32(2)$ for $(3,1)$, $\beta_1 = 0.33(1)$ for $(4,1)$, and $\beta_1 = 0.33(3)$ for $(5,1)$. Within error bars, $\beta_1$ for $l=1$ and different $k$ are equal. At $l=2$, we get $\beta_1 = 0.05(2)$ for $(3,2)$ and $\beta_1 = 0.04(2)$ for $(4,2)$. The exponents $\beta_1$ of $l=2$ are different from that of $l=1$. At $l=3$, $\beta_1 = 0.01(1)$ for $(4,3)$ and is different from that of $l=1$ and $l=2$. So the universality class of $(k,l)$ clique percolation is characterized by $l$.
In Fig.\ref{fig_phi}(b), the log-log plot of $\chi_\Delta$ versus $N$ is shown for different $(k,l)$. The slope of curve gives the exponent $\beta_2$.
It is found that $\beta_2$ is equal to the corresponding $\beta_1$ within error bars.

\begin{figure}[t!]
\centerline{\includegraphics[angle=0,width=0.75\columnwidth]{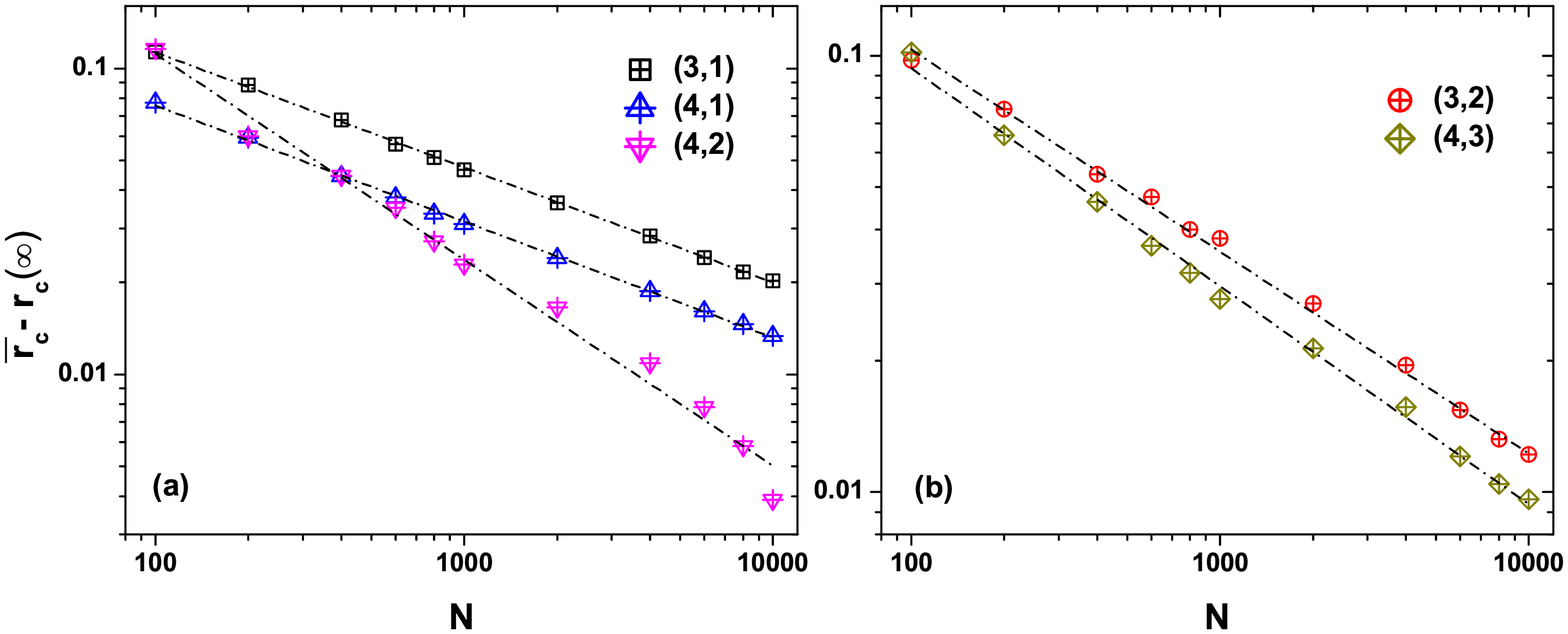}}
\caption{(color online) Log-log plot of $\bar{r}_c - r_c(\infty)$ versus $N$. The results of $l < k-1$ is shown in (a) and $l=k-1$ in (b).  The slopes of fitting lines give $1/\nu_{1}$, which is defined in Eq.~\ref{9} and summarized in Table I.
}
\label{fig_psi}
\end{figure}

The finite-size effects of $\bar{r}_c (N)- r_c (\infty)$ are shown in Fig.~\ref{fig_psi}(a) for $l < k-1$ and Fig.~\ref{fig_psi}(b) for $l=k-1$.
From the Monte Carlo data, the values of $r_c (\infty)$ and $\nu_1$ are determined simultaneously. For $l=k-1$, we get $1/\nu_1=0.46(2)$ for $(3,2)$
and $1/\nu_1=0.49(4)$ for $(4,3)$. For $l < k-1$, we obtain $1/\nu_1=0.37(2)$ for $(3,1)$,  $1/\nu_1=0.38(2)$ for $(4,1)$, and $1/\nu_1=0.67(20)$ for $(4,2)$. The $\bar{r}_c (N)$ of $(4,2)$ clique percolation shows deviation from a simple power law of $N$. This deviation could be resulted by additional finite-size terms. $1/\nu_1$ of different $k$ and the same $l$ are equal within error bars. We summarize $r_c (\infty)$ and $1/\nu_1$ of different $(k,l)$ clique percolation in Table I.

\begin{figure}[t!]
\centerline{\includegraphics[angle=0,width=0.75\columnwidth]{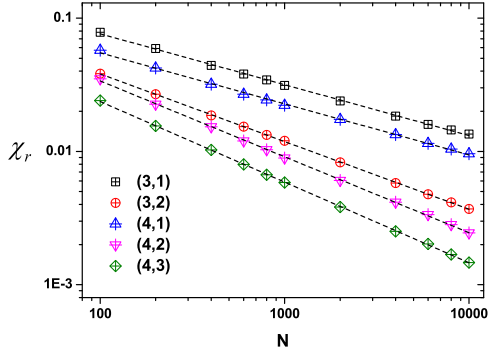}}
\caption{(color online) Log-log plot of $\chi_r$ versus $N$. The slopes of fitting lines give $1/\nu_2$, which is defined in Eq. \ref{10} and summarized in Table I.
}
\label{fig_psi2}
\end{figure}

The root-mean-square of $\delta r_c$ is shown in Fig.\ref{fig_psi2}. Our results of simulations confirm the power-law behavior of $\chi_r$. For $k=3$,
we get $1/\nu_2=0.33(5)$ for $l=1$ and $1/\nu_2=0.50(2)$ for $l=2$. At $k=4$, we obtain $1/\nu_2=0.35(3)$ for $l=1$, $1/\nu_2=0.52(4)$ for $l=2$,
and $1/\nu_2=0.60(3)$ for $l=3$. Other results of $1/\nu_2$ are given in Table I. $1/\nu_2$ of different $k$ and the same $l$
are equal within error bars. This confirms further that the universality of $(k,l)$ clique percolation is characterized alone by $l$. The exponent $\nu_2$ agrees with $\nu_1$ at $l=1,2$, but differs from $\nu_1$ at $l=3$.

\begin{figure}[t!]
\centerline{\includegraphics[angle=0,width=0.85\columnwidth]{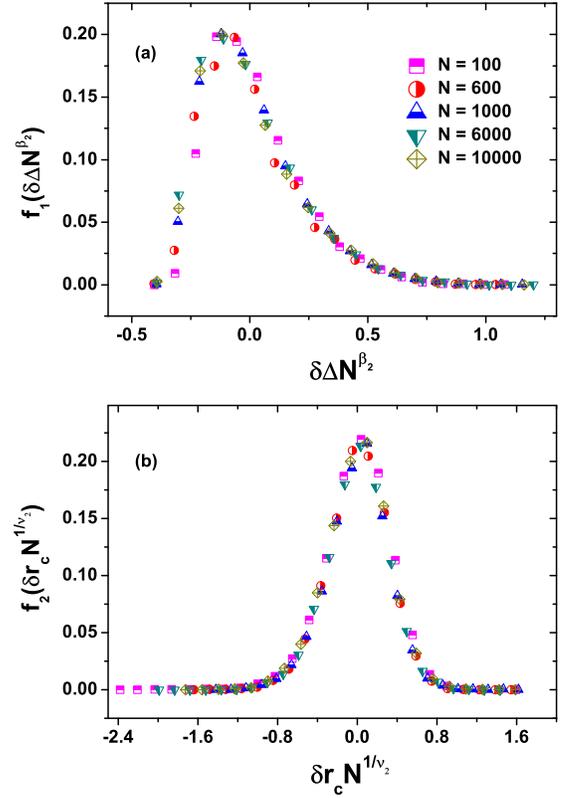}}
\caption{(color online) Finite-size scaling functions of fluctuation distribution functions (a) $P_\Delta (\delta \Delta, N)$ and (b) $P_r (\delta r_c, N)$ for $(4,1)$ clique percolation.
}
\label{fig_psci}
\end{figure}

For a network with size $N$, we simulate its evolution for many times. At each simulation, we can get the largest reduced size gap $\Delta$
and the corresponding transition point $r_c$. From the results of many simulations, the fluctuation distribution functions
$P_\Delta(\delta \Delta,N)$ and $P_r (\delta r_c,N)$ can be obtained. Five different network sizes $N = 100, 600, 1000, 6000, 10000$ are chosen.
Each fluctuation distribution function has five different curves. Using the finite-size scaling forms of Eq.~\ref{13} and \ref{14}, five curves of
each distribution function collapse into one curve of finite-size scaling function. The scaling variables are defined with the exponents $\beta_2$ and
 $\nu_2$ obtained from $\chi_\Delta$ and $\chi_r$. In Fig.~\ref{fig_psci}, the finite-size scaling functions $f_{1}(\delta \Delta N^{\beta_2})$
 and $f_{2}(\delta r_c N^{1/\nu_2})$ are shown for $(4,1)$ clique percolation. $f_1$ and $f_2$ of other $(k,l)$ clique percolation  have similar behavior.

In the evolution process of the ER model beginning from $N$ isolated nodes to being fully connected with $N(N-1)/2$ vertices, there are a series of $(k,l)$ clique percolation. The transition point of a $(k,l)$ clique percolation is $T_c = \frac{1}{2}a(k,l) N^{2-{2\over k+l-1}}$, which is of the same order of $N$ for different clique percolation with equal $k+1$. The sequence of clique percolation in the ER model is as following:
a) $(2,1)$ clique percolation at $T_c =0.5 N$, b) $(3,1)$ clique percolation at $T_c=0.5 N^{4/3}$, c) $(3,2)$ clique percolation
at $T_c =2^{-3/2} N^{3/2}\simeq 0.354 N^{3/2}$, d) $(4,1)$ clique percolation at $T_c =2^{-5/6} N^{3/2}\simeq 0.561 N^{3/2}$,
e) $(4,2)$ clique percolation at $T_c =2^{-1}2.5^{-1/5} N^{8/5}\simeq 0.416 N^{8/5}$, f) $(5,1)$ clique percolation
at $T_c =2^{-1} 6^{1/10} N^{8/5}\simeq 0.598 N^{8/5}$, g) $(4,3)$ clique percolation at $T_c =2^{-1}3^{-1/3} N^{5/3}\simeq 0.347 N^{5/3}$, and so on.

\begin{table}[htbp]
\centering  %
\caption{\label{comparison} Summary of reduced transition points $r_c  = T_c/N^{2-{2\over k+l-1}}$ and critical exponents of $(k,l)$ clique percolation.
$r_c(\infty)$ is obtained from Eq.~\ref{9} and the analytic result $r^{T}_c(\infty)$ is given in Eq.~\ref{9c}. The results of $(2,1)$ clique percolation
are taken from\cite{fanjingfang}.}
\begin{tabular}{c|c|c|c|c|c|c}
\hline
(k,l)   &$r_c(\infty)$   &$r^{T}_c(\infty)$   &$\beta_{1}$      &$\beta_{2}$       &$1/\nu_{1}$     &$1/\nu_{2}$         \\ \hline
(2,1)   &0.5006(7)       & $2^{-1}$      &0.331(5) &0.334(2)    &0.331(7)   &0.334(3) \\   %
(3,1)   &0.50(1)         & $2^{-1}$       &0.32(2)  &0.36(4)     &0.37(2)    &0.33(5)  \\   %
(3,2)   &0.34(2)         &$2^{-3/2}$  &0.05(2)  &0.03(2)     &0.46(2)    &0.50(2)  \\
(4,1)   &0.56(1)         &$2^{-5/6}$  &0.33(1)  &0.35(2)     &0.38(2)    &0.35(3)  \\
(4,2)   &0.40(1)         &$2^{-1}2.5^{-1/5}$ &0.04(2) &0.03(2)   &0.67(20)    &0.52(4)  \\
(4,3)   &0.34(2)         &$2^{-1}3^{-1/3}$ &0.01(1)   &0.01(1)   &0.49(4)    &0.60(3)  \\
(5,1)   &0.60(1)         &$2^{-1}6^{1/10}$   &0.33(3) &0.38(4) &0.30(5) &0.37(6)\\ \hline
\end{tabular}
\end{table}

For comparison with the investigations above, we investigate directly the size of the largest $(k,l)$ clique community in a network to study general clique percolation. For a network with $N$ vertices and $T$ edges, the reduced size of the largest $(k,l)$ clique community is denoted as $s_1 (T,N)$. Near the transition point $T_c$ of a $(k,l)$ clique percolation, we anticipate a finite-size scaling form
\begin{equation}
s_1(T,N) = N^{-\beta/\nu}\widetilde{s}_{1}(tN^{1/\nu}),
\label{15}
\end{equation}
where $t = (T-T_c)/T_c$ and $\nu$ is the critical exponent of correlation length. In the limit $N \to \infty$, $s_1 (T,\infty)=0$ at $T < T_c$
and $s_1 (T,\infty)= A_1 t^\beta$ at $T > T_c$. Our previous investigations \cite{fan2012,liu2012} has confirmed the finite-size scaling form Eq.~\ref{15} for $(2,1)$ clique percolation.

At the transition point $T_c$, $\left. s_1 (T,N)N^{\beta/\nu}\right|_{T=T_c}=\widetilde{s}_1 (0)$ and is independent of network size $N$.
Using this property, we can determine the transition point $T_c$ of $(k,l)$ clique percolation from the fixed point of $s_1 (T,N)N^{\beta/\nu}$.

\begin{figure}[t!]
\centerline{\includegraphics[angle=0,width=0.85\columnwidth]{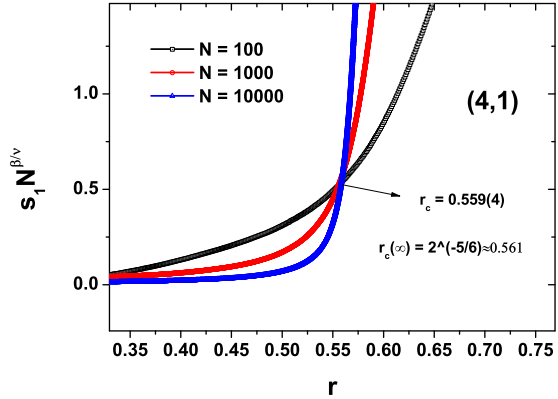}}
\caption{(color online) Scaling size of the largest $(4,1)$ clique community $s_1 N^{\beta/\nu}$ versus the reduced edge
number $r = T/N^{3/2}$. We take $\beta/\nu = 0.33$. There is a fixed point at $r_c = 0.559(4)$.
}
\label{fig_fixpoint}
\end{figure}

In Fig.~\ref{fig_fixpoint}, we plot $s_1 (T,N)N^{\beta/\nu}$ of $(4,1)$ clique community as a function of $r$ for different $N$. $\beta/\nu=0.33$ has been taken. We find a fixed point at $r_c=0.559(4)$,  which agrees with $r_c(\infty) = 0.56(1)$ obtained from network evolution. We expect that $\beta/\nu$ is equal to $\beta_2$ \cite{fanjingfang}.

In summary, we introduce a $(k,l)$ clique community consisting of adjacent $k$-cliques sharing at least $l$ vertices with $k-1 \ge l \ge 1$. There is a $(k,l)$ clique percolation with the emergence of a giant $(k,l)$ clique community. We study the $(k,l)$ clique percolation by investigating the
largest reduced size gap $\Delta$ of the largest clique community during network evolution and the corresponding evolution step $T_c$.
If the average size gap $\bar{\Delta} \sim N^{-\beta_1}$, there is a continuous clique percolation for $0 < \beta_1 < 1$ and a discontinuous clique percolation when $\beta_1 = 0$. The reduced transition point $\bar{r}_c$ obtained from $T_c$ has the finite-size effect $\bar{r}_c - r_c (\infty) \sim  N^{-1/\nu_1}$. The values of $r_c (\infty)$ obtained from Monte Carlo data agree with the analytic result of Eq.~\ref{8}, which is derived using generating function method. The sequence of $(k,l)$ clique percolation in the ER model is as following: $(2,1), (3,1), (3,2), (4,1), (4,2), (5,1), (4,3)$, and so on. The root-mean-squares of fluctuations $\delta \Delta$ and $\delta r_c$ are found to decay algebraically as $\chi_\Delta \sim N^{-\beta_2}$ and $\chi_r \sim N^{-1/\nu_2}$. The fluctuation distribution functions follow the finite-size scaling forms $P_\Delta(\delta \Delta,N) =  N^{\beta_{2}} f_{1}(\delta \Delta N^{\beta_2})$ and $P_r(\delta r_c,N) =  N^{1/\nu_{2}}f_{2}(\delta r_c N^{1/\nu_2})$, respectively.

The universality of $(k,l)$ clique percolation is characterized by the critical exponents $\beta_1$, $\beta_2$, $\nu_1$ and $\nu_2$.
It has been found that the critical exponents of $(k,l)$ clique percolation are independent of $k$, but dependent on $l$.
The universality class of $(k,l)$ clique percolation is characterized alone by $l$.

\begin{acknowledgments}
This work is supported by the National Natural Science Foundation
of China under grant 11121403.
\end{acknowledgments}

\end{document}